\DeclareRobustCommand{\VAN}[3]{#2}
\let\VANthebibliography\thebibliography
\def\thebibliography{\DeclareRobustCommand{\VAN}[3]{##3}\VANthebibliography}

\documentclass[fleqn,usenatbib]{mnras}
\usepackage{graphicx}
\usepackage{amsmath}	
\usepackage{amssymb}
\usepackage{newtxtext,newtxmath}
\usepackage{amsmath,amssymb,bm,hyperref,subfig,booktabs,multirow}
\usepackage{textcomp}
\usepackage[T1]{fontenc}
\usepackage{gensymb}
\usepackage[euler]{textgreek}
\usepackage{multicol,lipsum, babel}
\usepackage{natbib}
\usepackage{multirow}
\bibpunct{(}{)}{;}{a}{}{,} 
\usepackage{comment}
\hypersetup{colorlinks=true,citecolor=blue,
linkcolor=blue,filecolor=magenta,urlcolor=blue}

\title[RGB population in the foreground sub-structure of the SMC]{Presence of red giant population in the foreground stellar sub-structure of the Small Magellanic Cloud}

\author[James et al.]{
Dizna  James,$^{1}$\thanks{diznajames28@gmail.com}
Smitha  Subramanian,$^{2}$\thanks{smitha.subramanian@iiap.res.in} 
Abinaya O. Omkumar,$^{2,3,4}$
Adhya Mary,$^{1}$
Kenji Bekki,$^{5}$
\newauthor Maria-Rosa L. Cioni,$^{3}$
Richard de Grijs,$^{6,7}$
Dalal El Youssoufi,$^{3,4}$
Sreeja S. Kartha,$^{1}$
Florian Niederhofer,$^{3}$
\newauthor Jacco Th. van Loon$^{8}$
\\
$^{1}$Department of Physics, Christ (Deemed to be) University, Bangalore, India \\
$^{2}$Indian Institute of Astrophysics, Koramangala II Block, Bangalore, India\\
$^{3}$Leibniz-Institut f\"ur Astrophysik Potsdam (AIP), An der Sternwarte 16. D-14482 Potsdam, Germany\\
$^{4}$Institut f\"{u}r Physik und Astronomie, Universit\"{a}t Potsdam, Haus 28, Karl-Liebknecht-Str. 24/25, D-14476 Golm (Potsdam), Germany \\
$^{5}$ICRAR, M468, The University of Western Australia 35 Stirling Highway,Crawley Western Australia, 6009, Australia\\
$^{6}$Department of Physics and Astronomy, Macquarie University, Balaclava Road, Sydney, NSW 2109, Australia\\
$^{7}$Research Centre for Astronomy, Astrophysics and Astrophotonics, Macquarie University, Balaclava Road, Sydney, NSW 2109, Australia\\
$^{8}$ Lennard-Jones Laboratories, Keele University, ST5 5BG, UK\\
}

\date{Accepted XXX. Received YYY; in original form ZZZ}

\pubyear{2021}

\begin{document}
\label{firstpage}
\pagerange{\pageref{firstpage}--\pageref{lastpage}}
\maketitle

\begin{abstract}

The eastern region of the Small Magellanic Cloud (SMC) is found to have a foreground stellar sub-structure, which is identified as a distance bimodality ($\sim$ 12 kpc apart) in the previous studies using Red Clump (RC) stars. Interestingly, studies of Red Giant Branch (RGB) stars in the eastern SMC indicate a bimodal radial velocity (RV) distribution. In this study, we investigate the connection between these two bimodal distributions to better understand the nature and origin of the foreground stellar sub-structure in the eastern SMC. We use the \textit{Gaia} EDR3 astrometric data and archival RV data of RGB stars for this study. We found a bimodal RV distribution of RGB stars (separated by $\sim$ 35--45 km s$^{-1}$) in the eastern and south-western (SW) outer regions. The observed proper motion values of the lower and higher RV RGB components in the eastern regions are similar to those of the foreground and main-body RC stars respectively. This suggests that the two RGB populations in the eastern region are separated by a similar distance as those of the RC stars, and the RGB stars in the lower RV component are part of the foreground sub-structure. Based on the differences in the distance and RV of the two components, we estimated an approximate time of formation of this sub-structure as 307 $\pm$ 65 Myr ago. This is comparable with the values predicted by simulations for the recent epoch of tidal interaction between the Magellanic Clouds. Comparison of the observed properties of RGB stars, in the outer SW region, with \textit{N}-body simulations shows that the higher RV component in the SW region is at a farther distance than the main body, indicating the presence of a stellar Counter-Bridge in the SW region of the SMC.
\end{abstract}




\section{Introduction}
{\indent According to the $\Lambda$ cold dark matter model, galaxies grow in mass through the hierarchical assembly of smaller systems \citep{Fall1980,vandenbosch2002,Agertz2011}. Thus interactions and mergers play a significant role in the evolution of galaxies. One of the nearest examples of an on-going hierarchical merging process, is the Large Magellanic Cloud (LMC) -- Small Magellanic Cloud (SMC) galaxy pair, interacting mutually and with the Milky Way. The LMC and the SMC are two gas-rich interacting dwarf galaxies located at a distance of 50$\pm$2 kpc (LMC -- \citealp{degrijs2014lmc}) and 62$\pm$1 kpc (SMC -- \citealp{degrijs2015smc}) respectively. There exists a bridge of gas and stars connecting these galaxies known as the Magellanic Bridge (MB), a leading and a trailing stream of gas known as the Leading Arm (LA) and the Magellanic Stream (MS), respectively.  The MB, MS and the LA are prominent features in \ion{H}{i} maps \citep{putman2003}.\\ 
\indent Simulations of the Magellanic System  \citep{Besla2012,Diaz2012}, based on revised proper motion estimates of the Magellanic Clouds (MCs) (\citealp{kallivayalil2006,vieira2010}),  explain the formation of the observed gaseous features around the MCs as a result of their mutual interactions. The dominant nature of these interactions are suggested to be tidal, predicting the presence of stellar sub-structures along with the gaseous features around the MCs. According to these simulations, the MS and the MB were formed $\sim$ 1.5 Gyr and $\sim$ 100--300 Myr ago respectively,  mainly from  material stripped from the SMC. However, the MS has also been suggested to contain material stripped from the LMC (\citealp{nidever2008, hammer2015, Ritcher2017}) and based on the observed low metal abundance of stars in the MB, \cite{Rama} suggested that the time of the initial formation of the MB may date back to several billion years. Based on the relative motions of the MCs and recent proper motion measurements of stars within the MB region \citep{gaiadr1,gaia2018}, the tidal interaction event which formed the MB is suggested to have happened $\sim$ 150 Myr ago 
(\citealp{zivick2018,zivick2019,schmidt2019}).
Though tidal interactions must have played a dominant role, ram-pressure effects due to the Milky Way's halo could have also altered the present shape of the gaseous features of the Magellanic System   (\citealp{hammer2015,salem2015, Tepper2019,Wang2019}).\\ 
\indent Stellar sub-structures formed during the formation of the MB and MS are expected to contain stars older than $\sim$ 300 Myr and $\sim$ 1.5 Gyr respectively. Though several studies \citep{nidever2011,bagheri2013,noel2013,skowron2014,noel2015,belokurov2017,jacyszyn2017,jacyszyn2019, Massana2020} found intermediate-age/old (age $>$ 2 Gyr) stars around the MB region, the interpretations of their origin differed. While \cite{noel2013,noel2015} and \cite{carrera2017}  supported a tidal origin for these intermediate-age stars, \cite{jacyszyn2017} and \cite{wagner2017} suggested them as part of the overlapping stellar halos of the MCs. No conclusive evidence of a stellar counterpart (consisting of stars older than 1.5 Gyr) to the MS has been found so far. 
However, the MB and LA host stellar populations of few Myr (\citealp{demers1998,harris2007,chen2014,skowron2014,princewhelan2019, nidever2019} and references therein),  which might have formed from the gas stripped during the interactions of the MCs.\\ 
\indent \citet{Nidever2013} identified a foreground population ($\sim$ 10--12 kpc in front of the SMC main body) of red clump (RC) stars (which are standard candles) in four distinct 0.36 deg$^2$ fields at a radius of 4$^\circ$ from the SMC centre to the East (in the direction of the MB and the LMC). \cite{Subramanian2017} and 
\cite{tatton2020} studied this feature using a subset and a complete set of near-infrared data of RC stars, obtained from the VISTA (Visible and Infrared Survey Telescope for Astronomy) survey of the MCs (VMC; \citealp{cioni2011}), respectively. VMC data being continuous and homogeneous, allowed both these studies to trace this foreground RC feature over 2$\rlap{.}^{\circ}$5--4$\rlap{.}^{\circ}$0 from the SMC centre to the East. All these studies suggested that this foreground RC population could be the result of interactions between the MCs. \citet{Nidever2013} and \cite{Subramanian2017} suggested a tidal origin for this feature, a tidally stripped stellar population from the SMC during the interaction between the MCs around 300 Myr ago, which in turn might have caused the formation of the MB. However, \cite{tatton2020} added that as some of these sub-structures are traced
only by RC stars and not by RR Lyrae stars, tidal effects cannot fully explain this RC feature and ram pressure effects might also be involved.  \citet{Nidever2013}, \citet{Subramanian2017} and \cite{tatton2020} could not probe this feature beyond 4$\rlap{.}^{\circ}$0 from the SMC centre, due to the limited spatial coverage of the data. \\ 
\indent \cite{omkumar} analysed this dual RC feature using \textit{Gaia} data release 2 (DR2; \citealp{gaia2018}) data of a $\sim$314 deg$^2$ region centred on the SMC, which cover the entire SMC and a significant portion of the MB. They found that the \textit{Gaia} DR2 data trace this foreground RC feature from $\sim$ 2$\rlap{.}^{\circ}$5 to 5--6$^\circ$ from the optical centre of the SMC in the eastern regions, towards the MB and that it does not fully overlap with the MB in the plane of the sky. However, \cite{El_Youssoufi_2021} 
detected this feature out to a 10$^{\circ}$ distance from the SMC centre, in the direction of the MB. They used the NIR data from VISTA Hemisphere Survey (VHS, \citealt{mcmohan}). \cite{omkumar} found that the proper motion values of the bright (foreground) RC was found to be significantly larger than that of the faint (main body) RC, which is expected if  the bright RC is at a closer distance. 
The authors found that the foreground stellar population is kinematically distinct from the stellar population of the main body with $\sim$ 35 km s$^{-1}$ slower tangential velocity and moving to the North-West relative to the main body. They found that though the observed properties of this foreground RC feature are not completely consistent with those predicted by the simulations of \cite{Diaz2012}, a comparison indicated that the foreground stellar sub-structure is most likely a tidally stripped counterpart of the gaseous MB and might have formed from the inner disc (dominated by stars) of the SMC.\\ 
\indent A foreground stellar sub-structure formed due to the tidal interactions of the MCs  is expected to have all types of stars older than the epoch of interaction, which presumably formed this sub-structure. So if the foreground RC feature is formed due to tidal effects operating during the formation of the MB, $\sim$ 100 -- 300 Myr ago, then we expect SMC stars older than 300 Myr to have a line of sight distance distribution similar to what is observed for RC stars. In a study of RR Lyrae stars (older than 10 Gyr) in the SMC using VMC data, \cite{muraveva2018} did not find a clear distance bimodality in the 2$\rlap{.}^{\circ}$5--4$^\circ$ eastern regions (especially in the North-Eastern regions) as observed in the RC distribution. However, as suggested by \cite{omkumar}, if the foreground RC population is tidally stripped from the disc of the SMC and the RR Lyrae stars are located in the less disturbed spheroidal component then the observed difference in the line of sight distance distribution of RC stars and RR Lyrae stars in the SMC can be explained.\\
\indent Another numerous and homogeneously distributed intermediate-age/old stellar tracer in a galaxy are Red Giant Branch (RGB) stars. 
 The dominant population of the RGB stars in the SMC have an age of $\sim$ 5 Gyr (\citealt{Rubele2018}; \citealt{elyousaffi2019}). As RGB stars are not standard candles, it is not trivial to obtain their line of sight distance distribution. However, the tip of the RGB is a well known standard candle and is used to estimate the distance to nearby galaxies. \cite{Groenewegen2019} estimated the distance to different regions of the SMC in the VMC footprint and the results indicated a shorter distance in the eastern regions. A spectroscopic study of RGB stars in the central 4$^\circ$ region of the SMC by \cite{dobbie2014} identified a bimodality in their radial velocity distribution  in the eastern SMC, the regions where the RC distance bimodality is observed. They suggested that the bimodal radial velocity distribution could be due to the tidal interaction between the MCs.  It will be interesting to infer the distances to these RGB stars with bimodal radial velocity distribution and compare with the distance of the foreground RC feature. 
 Due to the difficulty in estimating the distances to these RGB stars, another way is to check the proper motion values of the RGB stars in the two components of the radial velocity distribution and compare it with the observed proper motion values of the foreground and the main body RC stars by \cite{omkumar}. If the RGB stars in the two radial velocity components are located at two different distances along the line of sight (similar to the foreground and the main body RC stars) then their relative proper motion value is expected to match that of the relative proper motion between the foreground and main body RC stars. 
 In this context, we combine the radial velocity data from \cite{dobbie2014} and \cite{deleo2020} with the proper motion data from the Early Data Release 3 (EDR3) of \textit{Gaia} to obtain 3D kinematic information of RGB stars in the central 4$\rlap{.}^{\circ}$5 
 radial region of the SMC and to understand the effect of tidal interaction on an another intermediate-age stellar tracer in the SMC. \\
\indent The structure of this paper is as follows. Data used in the study are described in Section \ref{section2}. The analysis and results are presented in Section \ref{section3}. Section \ref{section4} provides a summary. 
 } 
\section{Data}
\label{section2}
We use the radial velocity data of 4172 RGB stars in the SMC provided by \cite{dobbie2014} and 1861 RGB stars by \cite{deleo2020}.
The two data sets are consistent with each other and the typical difference in the measurements of common stars (175 stars) in the two data sets is $\sim$ 1 km s$^{-1}$. 
The sample is distributed in an area of $\sim$ 37.5 deg$^2$ of the central SMC. 
The optical spectra of these RGB stars were obtained with the multi object optical spectrograph, 2dF/AAOmega instrument on the 3.9 m Anglo-Australian Telescope at Siding Spring Observatory, Australia. 
The proper motion data were taken from \textit{Gaia} EDR3 \citep{Gaiaedr3}. 
Out of 5859 sources (after removing duplicate sources) with radial velocity data, we could extract the proper motion values for 5750 sources (which are matched within 1 arcsec). Stars with no proper motion measurements in the RA, $\mu_{\alpha}$, and Dec, $\mu_{\delta}$, directions are removed from our further analysis. Moreover, we applied a cut to the proper motion values based on the expected range in their values ($-$3 mas yr$^{-1}$ $\le$  $\mu_{\alpha}$  $\le$ $+$3 mas yr$^{-1}$ and $-$3 mas yr$^{-1}$ $\le$  $\mu_{\delta}$ $\le$ $+$3 mas yr$^{-1}$) predicted by simulations \citep{Diaz2012} for the SMC main body and stellar tidal features around the SMC. This cut reduces the number of stars for further analysis to 5664.

\subsection{Division of sub-regions}
\label{section2.1}
\begin{figure}
    \centering
    \includegraphics[width=8cm,height=8cm]{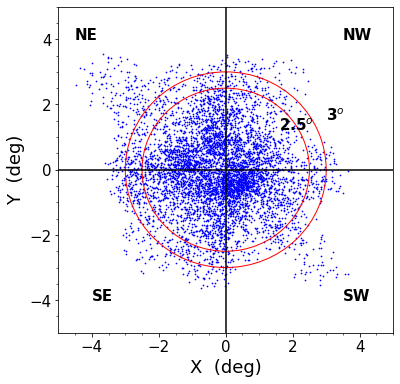}
    \caption{Cartesian plot showing the distribution of RGB sources (blue) analysed in this study. The red circle encompasses the 2$\rlap{.}^{\circ}$5 and 3$^{\circ}$ 
    radial region around the SMC from its optical centre \citep{deVaucouleurs1972} and the division of sectors are marked with the black lines.}
    \label{fig:xy_rgb}
\end{figure}

Fig. \ref{fig:xy_rgb} shows the spatial distribution of the RGB stars analysed in this study. The Cartesian plot (zenithal equidistant projection) is obtained by converting the RA and Dec of the RGB sources into X and Y respectively, using the equations provided by \cite{vandermarel2001} and considering the optical centre of the SMC  $\alpha_{S}$ = 00$^{h}$52$^m$12.5$^{s}$ and $\delta_{S}$ = $-72^{\circ}$49\arcmin43\arcsec (\citealt{deVaucouleurs1972}) as the origin. Initially the sources were divided into four sectors namely North East (NE, Y $>$ 0 \& X $<$ 0), North West (NW, Y $>$ 0 \& X $>$ 0), South East (SE, Y $<$ 0 \& X $<$ 0) and South West (SW, Y $<$ 0 \& X $>$ 0). Previous studies using the RC stars (\citealp{Subramanian2017,omkumar}) traced the foreground RC feature beyond 2$\rlap{.}^{\circ}$5 in the eastern regions of the SMC. Therefore each sector is further sub-divided into inner (\textit{R} $\leq$ 2$\rlap{.}^{\circ}$5) and outer (\textit{R} $>$ 2$\rlap{.}^{\circ}$5) sub-regions. The red circles in Fig. \ref{fig:xy_rgb} mark the 2$\rlap{.}^{\circ}$5 and 3$^{\circ}$ radial regions centred on the optical centre of the SMC, and the sector-wise division is shown using the black lines across the spatial distribution of the RGB sources.

\begin{figure*}
\centering
 \includegraphics[width=0.24\textwidth]{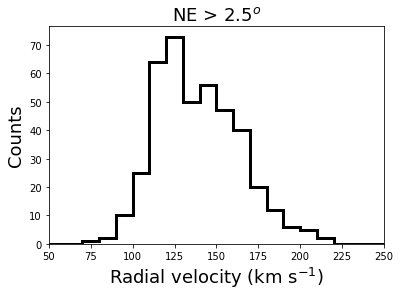}
 \includegraphics[width=0.24\textwidth]{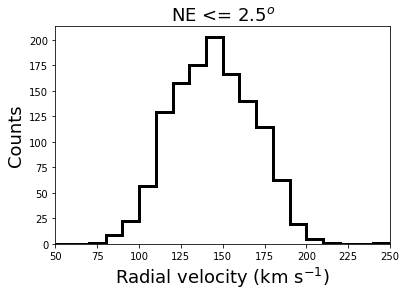}
 \includegraphics[width=0.24\textwidth]{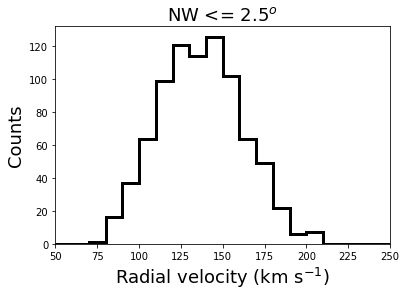}
 \includegraphics[width=0.24\textwidth]{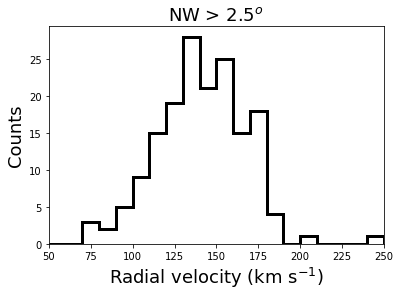}
 \includegraphics[width=0.24\textwidth]{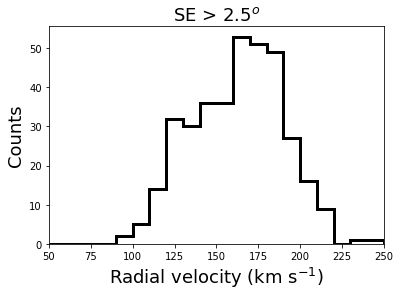}
 \includegraphics[width=0.24\textwidth]{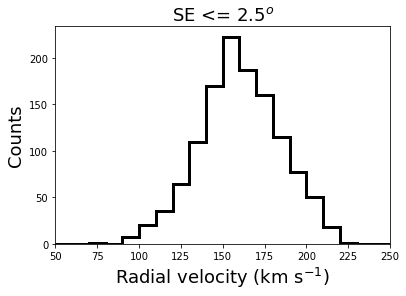}
 \includegraphics[width=0.24\textwidth]{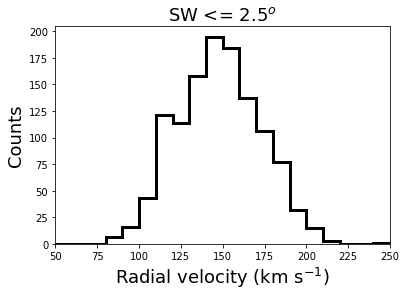}
 \includegraphics[width=0.24\textwidth]{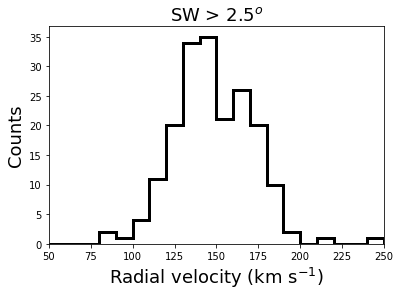}
\caption{Radial velocity distributions of northern and southern sub-regions are shown in the first and second rows, respectively. The first two panels in both rows represent eastern (outer; \textit{R} > 2$\rlap{.}^{\circ}$5  and inner; \textit{R} $\leq$ 2$\rlap{.}^{\circ}$5) sub-regions and the next two correspond to the western (inner; \textit{R} $\leq$ 2$\rlap{.}^{\circ}$5  and outer; \textit{R} > 2$\rlap{.}^{\circ}$5) sub-regions, respectively. 
}
\label{fig:2}
\end{figure*}

\begin{figure*}
    \centering
    
    \subfloat{\includegraphics[width=0.38\textwidth]{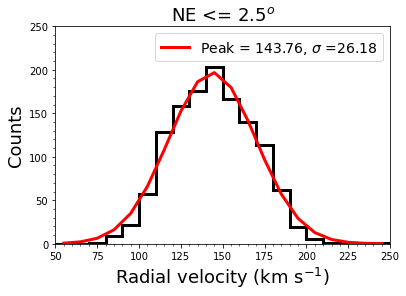}}
    \subfloat{\includegraphics[width=0.38\textwidth]{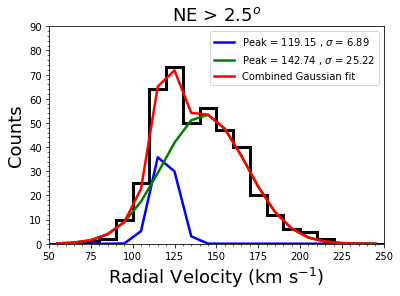}}\\
    \subfloat{\includegraphics[width=0.38\textwidth]{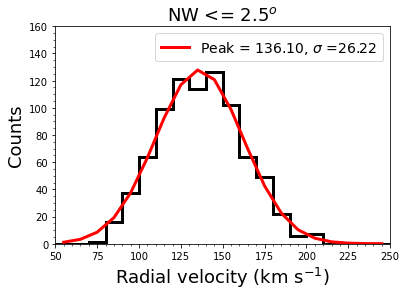}}
    \subfloat{\includegraphics[width=0.38\textwidth]{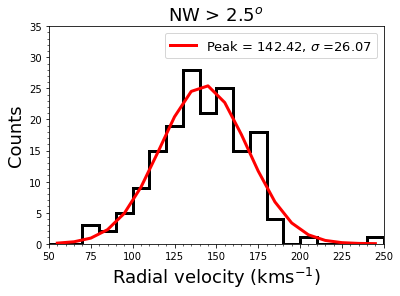}}\\
    \subfloat{\includegraphics[width=0.38\textwidth]{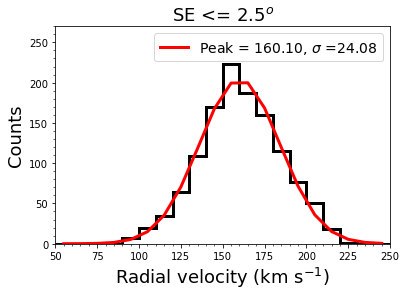}}
    \subfloat{\includegraphics[width=0.38\textwidth]{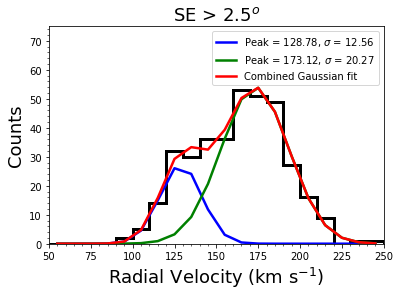}}\\
    \subfloat{\includegraphics[width=0.38\textwidth]{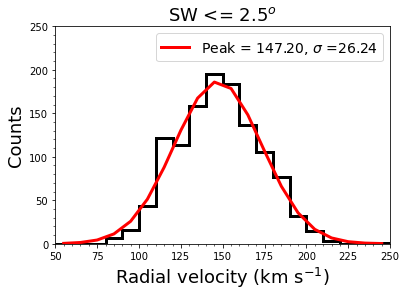}}
    \subfloat{\includegraphics[width=0.38\textwidth]{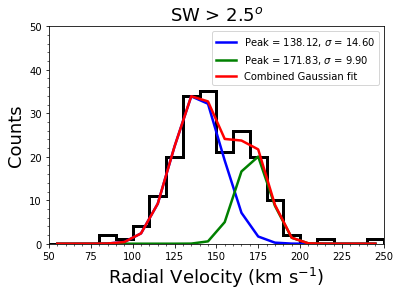}}\\
    
    \caption{Radial velocity distributions (black line) along with their best fitting Gaussian components are shown (blue and green lines) and the total fit is marked with a red line for the selected RGB sources in the NE (top row), NW (second row), SE (third row)  and SW (bottom row) inner (\textit{R} $\leq$ 2$\rlap{.}^{\circ}$5; left panel) and outer (\textit{R} $>$ 2$\rlap{.}^{\circ}$5; right panel) sub-regions, respectively.}
    \label{fig:3}
\end{figure*}
\begin{figure*}
    \centering
    \subfloat{\includegraphics[width=0.38\textwidth]{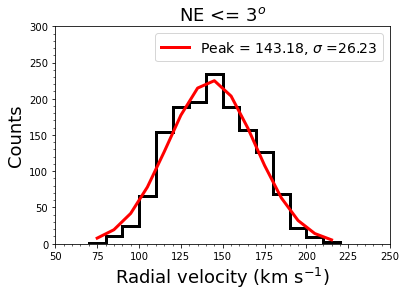}}
    \subfloat{\includegraphics[width=0.38\textwidth]{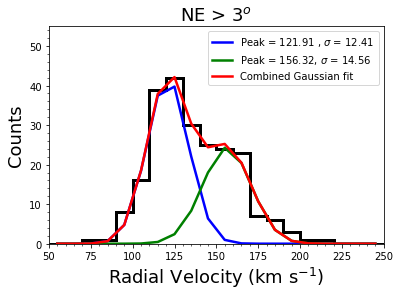}}\\
    \subfloat{\includegraphics[width=0.38\textwidth]{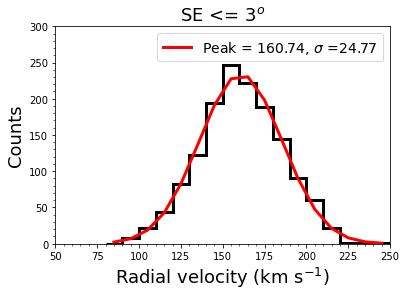}}
    \subfloat{\includegraphics[width=0.38\textwidth]{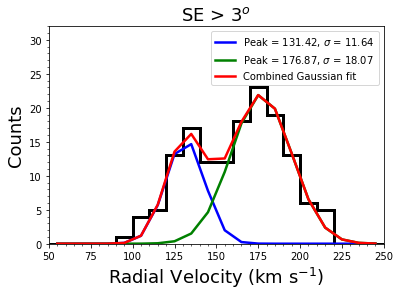}}\\
    
    \caption{Radial velocity distributions (black line) with radial cut \textit{R} $\leq 3^{\circ}$ and \textit{R} > $3^{\circ}$ along with their best fitting Gaussian components in the NE outer (top right) and SE outer (bottom right) regions are shown. The total fit is marked as in Fig. \ref{fig:3} for the selected RGB sources.}
    \label{fig:4}
\end{figure*}
\section{Analysis and Results}
\label{section3}
In this study, we analyse the radial velocities and the proper motions of the selected RGB stars. We compare the results from the RGB stars with the proper motion values of the RC stars. We also compare our results with simulations of the SMC \citep{Diaz2012}.
\subsection{Radial velocity distribution of RGB stars}
\label{section 3.1}
The radial velocity distributions of each of the sub-regions are obtained. The Majority of the stars have  radial velocity values between 100 -- 200 km s$^{-1}$. The average uncertainty in the radial velocities is less than 5 km s$^{-1}$. The number of stars in different sub-regions (described in Section \ref{section2.1}) ranges from 148 to 1451.
Based on the number of stars in different sub-regions and the radial velocity range, we selected an optimal bin size of 10 km s$^{-1}$ (to have number of bins $\le$ (number of stars in each sub-region)$^{0.5}$) to obtain the radial velocity distribution of stars in sub-regions. 
Fig. \ref{fig:2} shows the radial velocity distribution of sub-regions and the panels are arranged according to the spatial distribution of the RGB stars (Fig. \ref{fig:xy_rgb}) on the sky.
The outer regions (\textit{R} $>$ 2$\rlap{.}^{\circ}$5) are relatively less populated compared with the inner regions. 
The outer NE (\textit{R} $>$ 2$\rlap{.}^{\circ}$5) and SE (\textit{R} $>$ 2$\rlap{.}^{\circ}$5) sectors show clear bi-modality (top left and bottom left panels of Fig. \ref{fig:2}). 
The observed radial velocity distributions are fitted with Gaussian functions to find the peak velocities. The observed distributions are initially fitted using a single Gaussian function. An additional Gaussian is added if the reduced $\chi^2$ of the fit improves by 30 per cent or more, compared to the reduced $\chi^2$ value of a single Gaussian fit and the full width at half maximum (FWHM) of the second Gaussian component is larger than the bin-size of the distribution. We use the python program curve-fit for the fitting procedure. The sub regions with their best fitting profiles are shown in Fig. \ref{fig:3}. The best-fitting parameters along with their errors and $\chi^2$ values are provided in Table \ref{tab:1}.\\ 
\begin{table*}
\caption{ Gaussian fit parameters for the radial velocity distributions of RGB stars.}
\label{tab:1}
\resizebox{\textwidth}{!}{
\begin{tabular}{cccccc}
\hline
                            &               & \multicolumn{2}{c}{Lower velocity RGB} & \multicolumn{2}{c}{Higher velocity RGB}                                                                        \\ \cline{3-6}
\multirow{-2}{*}{Regions} & \multirow{-2}{*}{Stellar Counts}                                           & Peak 1 (km s$^{-1}$)                  & Sigma 1 (km s$^{-1}$)                & Peak 2 (km s$^{-1}$)                    & Sigma 2 (km s$^{-1}$)                      \\ \hline
NE Outer (\textit{R} $>$ 2$\rlap{.}^{\circ}$5)&413                  & 119.15 ± 0.49                         & 6.89 ± 3.76                          & 142.74 ± 1.46                           & 25.22 ± 6.76                                                      \\
NE Inner  (\textit{R} $\leq 2\rlap{.}^{\circ}$5)&1262                & -                                     & -                                    & 143.76 ± 0.86                          & 26.18 ± 6.72                                                   \\
NW Inner   (\textit{R} $\leq 2\rlap{.}^{\circ}$5)&828                & -                                     & -                                    & 136.10 ± 0.70                           & 26.22 ± 6.08                                                  \\
NW Outer  (\textit{R} $>$ 2$\rlap{.}^{\circ}$5)&166                 & -                                     & -                                    & 142.42 ± 1.71                           & 26.07 ± 9.43                                                    \\
SE Outer  (\textit{R} $>$ 2$\rlap{.}^{\circ}$5)&362      & 128.78 ± 1.92                                     & 12.56 ± 6.50 & 173.12 ± 1.40         & 20.27 ± 7.40                            \\
SE Inner   (\textit{R} $\leq 2\rlap{.}^{\circ}$5)&1237                & -                                     & -                                    & 160.10 ± 0.69                          & 24.08 ± 5.77                                                     \\
SW Inner   (\textit{R} $\leq 2\rlap{.}^{\circ}$5)&1208                 & -                                     & -                                    & 147.20 ± 1.02                          & 26.24 ± 7.30                                                  \\
SW Outer  (\textit{R} $>$ 2$\rlap{.}^{\circ}$5)&188      & 138.12 ± 1.28                                    & 14.60 ± 5.90                         & 171.83 ± 1.58                          & 9.90 ± 5.14                                                    \\\hline
NE Outer (\textit{R} $>$ 3$^{\circ}$)&227           & 121.91 ± 1.65                        & 12.41 ± 5.64                        & 156.32 ± 3.22                           & 14.56 ± 8.31                                                  \\
NE Inner (\textit{R} $\leq 3^{\circ}$)&1448         & -                                     & -                                    & 143.18 ±  1.14 & 26.23  ±   7.76                          \\
SE outer (\textit{R} $>$ 3$^{\circ}$)&148           & 131.42 ± 1.25                         & 11.64 ±  5.27                        & 176.87 ±  1.11                          & 18.07 ± 6.54                                                  \\
SE Inner (\textit{R} $\leq 3^{\circ}$)&1451                & -                                     & -                                    & 160.74 ±  0.56  & 24.77 ±  5.28    \\\hline
\end{tabular}
}

\end{table*}

\indent The observed radial velocity distribution of RGB stars in all inner regions and NW outer region are well fitted with a single Gaussian component (Fig. \ref{fig:3}). 
We can see that the observed radial distributions of the outer NE and SE regions require two Gaussian components. The individual Gaussian components, the lower and higher velocity components (blue and green respectively), and the combined fit (red) are shown. This suggests a dual population of RGB stars in the NE and SE $\textit{R} > 2\rlap{.}^{\circ}5$. Though there are two components, they overlap significantly. The difference between the peaks of the two components is less than the sum of the widths of the two distributions. Since the components are not well separated, it is difficult to identify/select the probable members of each component. Hence we selected the inner and outer sub-regions in the NE and SE using the \textit{R} $\leq 3^{\circ}$ and \textit{R} $>$ $3^{\circ}$ criterion respectively. The radial velocity distributions showed clear bimodality in the outer sub-regions. The best fitting profile of the distribution of stars in the NE and SE outer sub-regions is a combination of two Gaussian functions which are well separated (Fig. \ref{fig:4}). The inner sub-regions are well fitted with a single Gaussian. The best fitting parameters for \textit{R} $\leq 3^{\circ}$ and \textit{R} $>$ $3^{\circ}$ in the NE and SE are also shown in Table \ref{tab:1}. The lower radial velocity component in the NE outer sub-region is found to have a larger number of stars than the higher radial velocity component and the trend is opposite in the SE outer sub-region. But note that the spectroscopic data used in our study could be incomplete and any comparison of the relative amplitudes of different components may not help to provide any reliable  scientific insights. We performed the entire analysis with bin sizes of 5 km s$^{-1}$ and 15 km s$^{-1}$ for the radial velocity distributions of all the sub-regions and found that a change in the bin-size does not affect our final results. For further analysis and estimation of proper motions, we use the values based on the \textit{R} = $3^{\circ}$ cut for the NE and SE regions.
Fig. \ref{fig:3} also shows two well separated velocity components in the SW  \textit{R} $>$ $2\rlap{.}^{\circ}5$ outer region  indicating the presence of a dual RGB population in the outer SW sub-region as well.\\
\indent The peak value of the higher velocity component of the NE (\textit{R} $>$ 3$^{\circ}$) and SE (\textit{R} $>$ 3$^{\circ}$) outer sub-regions is more similar/closer to the peak value of the single component in the respective inner regions. This suggests that the higher velocity component represents the radial velocity distribution of most of the stars in the SMC and hence could be the velocity distribution of stars in the main body of the SMC. The lower velocity component in the outer sub-regions might be the radial velocity distribution of the tidally affected stellar population. The difference in radial velocities between the two components are  $\sim$ 35 km s$^{-1}$ and $\sim$ 45 km s$^{-1}$ in the outer NE region (\textit{R} $>$ 3$^{\circ}$) and the outer SE (\textit{R} $>$ 3$^{\circ}$) sub-regions respectively. However, in the SW (\textit{R} $>$ 2$\rlap{.}^{\circ}$5) outer sub-region, the peak of the lower velocity component is similar to the peak of the single component in the inner SW sub-region. 
This indicates the stars in the higher velocity component in the SW outer sub-region could be associated with a sub-structure. 
The velocity difference between the two components in the outer SW region (\textit{R} $>$ 2$\rlap{.}^{\circ}$5) is $\sim$ 34 km s$^{-1}$. In order to better understand the connection between the different velocity components and the sub-structures we compare their proper motion values with those of the RC stars.\\
 
\begin{table*}
\caption{Median proper motion values of RGB and RC stars.}
\label{tab:2}
\resizebox{\textwidth}{!}{%
\begin{tabular}{ccccccccc}
\hline
\multirow{2}{*}{Regions} & \multicolumn{2}{c}{Lower Velocity RGB}                                    & \multicolumn{2}{c}{Higher Velocity RGB}                                   & \multicolumn{2}{c}{Foreground RC}                                         & \multicolumn{2}{c}{Main body RC}                                          \\ \cline{2-9} 
                         & \textit{μ$_\alpha$} (mas yr$^{-1}$) & \textit{μ$_\delta$} (mas yr$^{-1}$) & \textit{μ$_\alpha$} (mas yr$^{-1}$) & \textit{μ$_\delta$} (mas yr$^{-1}$)& \textit{μ$_\alpha$} (mas yr$^{-1}$) & \textit{μ$_\delta$} (mas yr$^{-1}$) & \textit{μ$_\alpha$} (mas yr$^{-1}$) & \textit{μ$_\delta$} (mas yr$^{-1}$) \\ \hline
NE Outer (\textit{R} $> 3^{\circ}$)              & 1.088 ± 0.021                       & --1.250 ± 0.014                      & 0.885 ± 0.029                       & --1.188 ± 0.017                      & 1.092 ± 0.007                       & --1.241±0.006                        & 0.902 ± 0.007                       & --1.161 ± 0.006                      \\
NE Inner (\textit{R} $\leq 3^{\circ}$)              & -                                   & -                                   & 0.780 ± 0.007                       & --1.217 ± 0.008                      & -                                   & -                                   & 0.795 ± 0.002                       & --1.167 ± 0.002                      \\
NW Inner (\textit{R} $\leq 2\rlap{.}^{\circ}$5)                 & -                                   & -                                   & 0.623 ± 0.008                       & --1.262 ± 0.006                      & -                                   & -                                   & 0.649 ± 0.003                       & --1.219 ± 0.002                     \\
NW Outer (\textit{R} $> 2\rlap{.}^{\circ}$5)                & -                                   & -                                   & 0.520 ± 0.017                       & --1.269 ± 0.018                      & -                                   & -                                   & 0.532 ± 0.006                       & --1.230 ± 0.006                      \\
SE Outer (\textit{R} $> 3^{\circ}$)              & 1.156 ± 0.038                       & --1.327 ± 0.025                      & 0.956 ± 0.027                       & --1.207 ± 0.022                      & 1.101 ± 0.008                       & --1.296 ± 0.007                      & 0.928 ± 0.008                       & --1.150 ± 0.007                       \\
SE Inner (\textit{R} $\leq 3^{\circ}$)              & -                                   & -                                   & 0.765 ± 0.007                       & --1.214 ± 0.007                      & -                                   & -                                   & 0.768 ± 0.002                         & --1.172 ± 0.002                        \\
SW Inner (\textit{R} $\leq 2\rlap{.}^{\circ}$5)                & -                                   & -                                   & 0.585 ± 0.007                       & --1.232 ± 0.006                      & -                                   & -                                   & 0.606 ± 0.002                       & --1.211 ± 0.002                     \\
SW Outer (\textit{R} $> 2\rlap{.}^{\circ}$5)                 & 0.489 ± 0.029                       & --1.304 ± 0.023                      & 0.460 ± 0.033                       & --1.257 ± 0.022                      & -                                   & -                                   & 0.471 ± 0.005                       & --1.240 ± 0.004                     \\ \hline
\end{tabular}%
}
\end{table*}

\begin{table*}
\centering
\caption{Magnitudes and distances of the RC stars in each sub-region.}
\label{tab:mag_dist}
\resizebox{\textwidth}{!}{%
\begin{tabular}{|c|c|c|c|c|c|c|}
\hline
                                                                   & \multicolumn{3}{c|}{Faint RC}                                 & \multicolumn{3}{c|}{Bright RC}                                \\ \hline
Sub-regions                                                        & Peak $\pm$ error   & Width $\pm$ error & Distance $\pm$ error & Peak $\pm$ error   & Width $\pm$ error & Distance $\pm$ error \\ \hline
NE Outer (\textit{R} \textbf{$>   3^{\circ}$})             & 18.932 $\pm$ 0.045 & 0.194 $\pm$ 0.031 & 66.375 $\pm$ 5.908   & 18.513 $\pm$ 0.046 & 0.145 $\pm$ 0.022 & 54.712 $\pm$ 3.639   \\ 
NE Inner (\textit{R} $\leq   3^{\circ}$)          & 18.778 $\pm$ 0.004 & 0.185 $\pm$ 0.004 & 61.808 $\pm$ 5.265   & -                  & -                 & -                    \\ 
NW Inner (\textit{R} $\leq   2\rlap{.}^{\circ}$5) & 18.781 $\pm$ 0.005 & 0.188 $\pm$ 0.006 & 61.913 $\pm$ 5.359   & -                  & -                 & -                    \\ 
NW Outer (\textit{R} $>   2\rlap{.}^{\circ}$5)    & 18.987 $\pm$ 0.008 & 0.173 $\pm$ 0.007 & 68.054 $\pm$ 5.413   & -                  & -                 & -                    \\ 
SE Outer (\textit{R} $>   3^{\circ}$)             & 18.981 $\pm$ 0.019 & 0.164 $\pm$ 0.014 & 67.866 $\pm$ 5.113   & 18.512 $\pm$ 0.036 & 0.141 $\pm$ 0.023 & 54.701 $\pm$ 3.554   \\ 
SE Inner (\textit{R} $\leq   3^{\circ}$)          & 18.795 $\pm$ 0.005 & 0.185 $\pm$ 0.005 & 62.292 $\pm$ 5.305   & -                  & -                 & -                    \\ 
SW Inner (\textit{R} $\leq   2\rlap{.}^{\circ}$5) & 18.831 $\pm$ 0.005 & 0.195 $\pm$ 0.005 & 63.344 $\pm$ 5.668   & -                  & -                 & -                    \\ 
SW Outer (\textit{R} $>   2\rlap{.}^{\circ}$5)   & 18.970 $\pm$ 0.009  & 0.172 $\pm$ 0.008 & 67.533 $\pm$ 5.344   & -                  & -                 & -                    \\ \hline
\end{tabular}%
}
\end{table*}

\begin{figure*}
    \centering
    \subfloat{\includegraphics[scale=0.4]{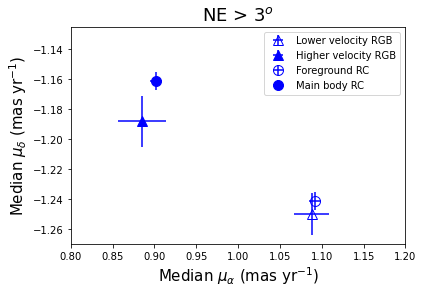}}
    \subfloat{\includegraphics[scale=0.4]{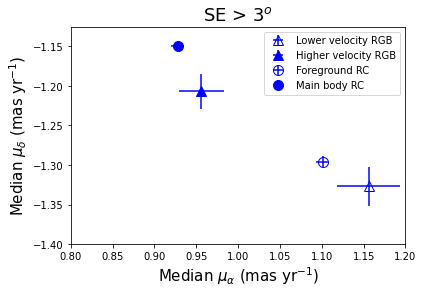}}
    \subfloat{\includegraphics[scale=0.4]{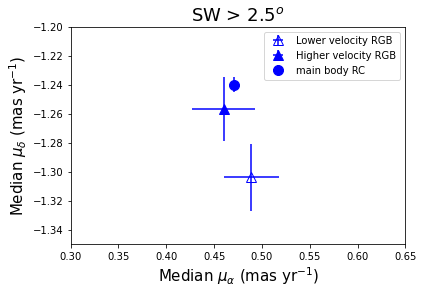}}\\
    \caption{Observed median proper motions for lower velocity RGB sources (open triangle) and higher velocity RGB sources (filled triangle) in the NE outer (\textit{R} $>$ $3^{\circ}$), SE outer(\textit{R} $>$ 3$^{\circ}$) and SW outer (\textit{R} $>$ 2$\rlap{.}^{\circ}$5) sub-regions. Observed median proper motions for foreground RC (open circle) and main body RC (filled circle) are shown for comparison. 
    }
    \label{fig:5}
\end{figure*}
\subsection{Comparison of the proper motion values of the RGB and RC stars}
\label{section 3.2}

To select the RGB stars corresponding to the single/double velocity components and analyse their proper motion properties in different sub-regions, we applied the following selection criterion. The stars which have velocities within the range, peak velocity $-$ $\sigma$ and peak velocity + $\sigma$, are selected and considered for further analysis. The peak velocity and the $\sigma$ are the best fitting values given in Table \ref{tab:1}. In the outer NE, outer SE and outer SW sub-regions, where dual RGB populations are found in the radial velocity distribution, the RGB stars in the lower and higher velocity components are separated using the respective peak velocity and $\sigma$ values of the respective components. For these selected RGB stars in different sub-regions, we estimated the median proper motion values in the RA and Dec directions. Before estimating the median values, we excluded stars having larger uncertainties ($>$ 0.1 mas yr$^{-1}$, which is two times the typical error associated with the proper motion values) in the proper motions. The  median \textit{$\mu_\alpha$} and \textit{$\mu_\delta$} along with their standard errors for the inner and outer regions are tabulated in Table \ref{tab:2}.

From Table \ref{tab:2}, we can see that the proper motion values of the lower velocity component, especially in the NE outer and SE outer regions are larger than the proper motion values of the higher velocity component. 
The obtained values of the two proper motion components of the lower and higher radial velocity components of RGB stars in the eastern sub-regions and their differences are comparable with the proper motion values of the foreground and main body population of the RC stars in the 2$\rlap{.}^{\circ}$5 -- 4$\rlap{.}^{\circ}$5 radial range and their differences as given by \cite{omkumar} (see their Tables 3, 4 and Figure 10) using \textit{Gaia} DR2.\\ 
\indent It will be interesting to make a comparison of the proper motions of the two radial velocity components of RGB stars with those of the two RC populations (foreground and main body) using the improved proper motion estimates from \textit{Gaia} EDR3. Hence, we obtained the 5$\rlap{.}^{\circ}$0 data of the SMC centred on the optical centre of the SMC by applying the same quality cuts used in \cite{omkumar} from \textit{Gaia} EDR3. 
We selected sources that are distributed in the same regions of the sky as in the studies of \cite{dobbie2014} and \cite{deleo2020} and merged them. The Gaia magnitudes  (\textit{G, G$_{BP}$, G$_{RP}$}) of the sources are corrected for interstellar extinction using the  extinction map ($\sim$ 3$^{\circ}$ radial region of the SMC) provided by \cite{Rubele2018}. For the stars in the region outside the coverage of their extinction map, we used the extinction values from the nearest region. We then divided the entire sample spatially to create similar sub-regions as mentioned in Section \ref{section2.1} and selected RC stars from the ${G_0}$ vs. (G$_{BP}$ - G$_{RP}$)$_0$ colour magnitude diagrams of the sub-regions. We produced the RC magnitude distributions in G$_{0}$ band with a bin size of 0.1 mag. The observed magnitude distributions of each sub-region is initially fit with a Gaussian function (for RC stars) and a quadratic polynomial (to account for RGB contamination in the RC selection) by using the curvefit function in PYTHON-SCIPY (\citealt{virtanen2020}) and the best-fit parameters (magnitudes and width along with their errors) are obtained.  
The fit to each of the sub-regions is repeated by adding an additional Gaussian component and further considered only if the $\chi^2$ of the fit is improved by 25 per cent than with a single Gaussian fit and the width of the second Gaussian is larger than the bin size of the distribution. Using this approach, we found that the NE and SE outer sub-regions have dual RC populations. The best fit parameters obtained for all the sub-regions are given in Table \ref{tab:mag_dist}. 
As described in Section 5 of \citep{omkumar}, we estimated the distances to each sub-region. In the eastern outer sub-regions, where dual RC populations are found, we calculated the distances corresponding to the faint and bright RC populations (faint RC corresponds to the main body of the SMC and bright RC corresponds to a foreground population, refer Table \ref{tab:mag_dist} for distance estimates). We then calculated the median proper motions (Table \ref{tab:2}) of the selected RC stars in each sub-region along with their respective uncertainties (which are standard errors associated with the median values) as described in Section 6 of \citep{omkumar}. The estimated differences in the $\mu_\alpha$ and $\mu_\delta$ values of the two RC populations in the NE outer (R $>$ 3$^{\circ}$) sub-region are $\sim$ 0.19 $\pm$ 0.01 mas yr$^{-1}$ and $\sim$ 0.08 $\pm$ 0.01 mas yr$^{-1}$, respectively. Similarly, the differences in the $\mu_\alpha$ and $\mu_\delta$ values in the SE outer (R $>$ 3$^{\circ}$)  sub-region are $\sim$ 0.17 $\pm$ 0.01 mas yr$^{-1}$ and $\sim$ 0.15 $\pm$ 0.01 mas yr$^{-1}$, respectively. We did not find any distance bimodality in the SW outer sub-region in our analysis using RC stars.\\

\indent Fig. \ref{fig:5} shows the median values of RGB stars in the lower and higher radial velocity components of the radial velocity distribution of the NE, SE and SW outer sub-regions in the \textit{$\mu_\alpha$} vs. \textit{$\mu_\delta$} plane. The median \textit{$\mu_\alpha$} and \textit{$\mu_\delta$} of the foreground and main body RC populations (in the NE and SE) are also shown in the left and middle panels of the plot for comparison. In the case of the SW outer sub-region, as only a single RC component with distance similar to that of the main body is found, the median \textit{$\mu_\alpha$} and \textit{$\mu_\delta$} of the main body RC is shown in the right panel of the plot. The difference in the proper motion values between the two RGB populations is evident for the NE and SE outer sub-regions. Interestingly, in these sub-regions the proper motion values of RGB stars in the lower velocity component are very similar or comparable within errors to those of the foreground RC stars. This suggests that the RGB stars with lower velocities in the NE and SE outer sub-regions are part of the foreground sub-structure identified using RC stars. The proper motion values of RGB stars in the higher velocity component do not exactly match with those of the main body RC population. The $\mu_{\alpha}$ values are very similar, but the $\mu_{\delta}$ values  of RGB stars are larger than those of the RC stars. A similar difference in $\mu_{\delta}$ values is observed between the single velocity component RGB and single component RC in all other sub-regions (see Table \ref{tab:2}). Since the discrepancy is only in one proper motion component, this rather points towards an intrinsic difference in velocity between the RGB and RC stars in the main body. If the RGB stars in the main body would be at a closer distance than the RC stars in the main body, both proper motion components would have been affected equally.\\ 
\indent In the SW outer sub-region we compare the proper motion values from the higher and lower velocity RGB components with the proper motion value of the single component RC (corresponding to the main body distance) found in this region. The difference between the lower and higher velocity components is very small, with negligible difference in $\mu_{\alpha}$. The proper motion value of the higher velocity RGB component seems to be more similar to that of the main body RC. However, as there is a negligible difference in the $\mu_{\alpha}$ values of both the lower and higher velocity components and there seems to exist an intrinsic difference in the $\mu_{\delta}$ values of the main body RGB and RC, it is difficult to say whether the lower or the higher velocity RGB component is associated with the main body.\\
\indent In the next sub-section we compare the observed radial velocity and proper motion values of different RGB components with the values predicted for the SMC main body and the associated tidal features, from the \textit{N}-body simulations \citep{Diaz2012}. 

\begin{figure*}
    \centering
    \subfloat{\includegraphics[scale=0.32]{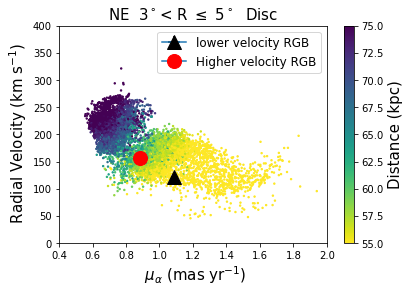}}
    \subfloat{\includegraphics[scale=0.32]{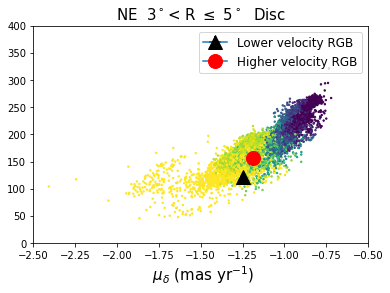}}
    \subfloat{\includegraphics[scale=0.32]{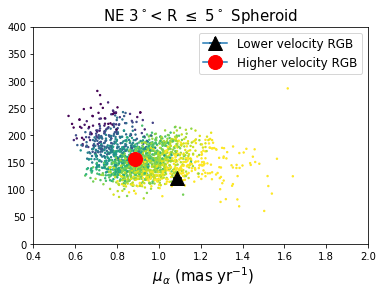}}
    \subfloat{\includegraphics[scale=0.32]{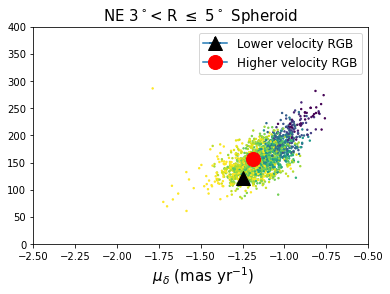}}\\
   
    \subfloat{\includegraphics[scale=0.32]{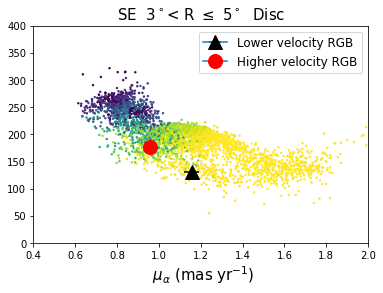}}
    \subfloat{\includegraphics[scale=0.32]{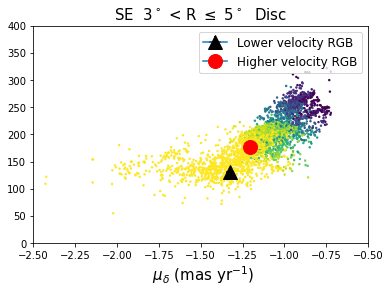}}
    \subfloat{\includegraphics[scale=0.32]{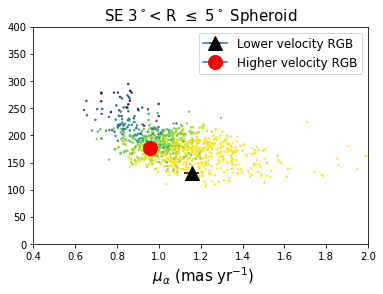}}
    \subfloat{\includegraphics[scale=0.32]{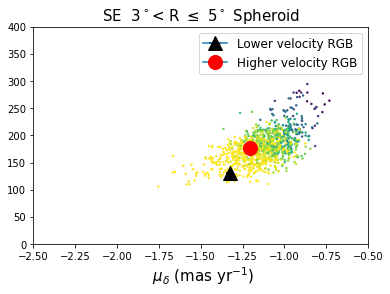}}\\
   
    \subfloat{\includegraphics[scale=0.32]{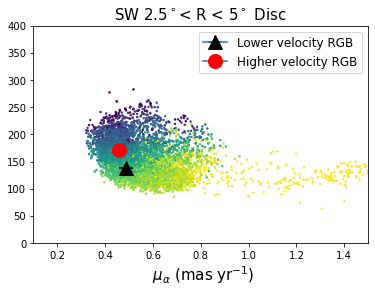}}
    \subfloat{\includegraphics[scale=0.32]{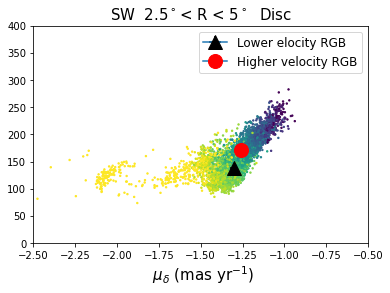}}
    \subfloat{\includegraphics[scale=0.32]{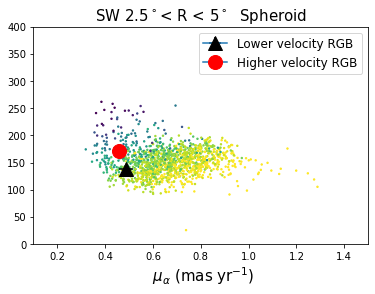}}
    \subfloat{\includegraphics[scale=0.32]{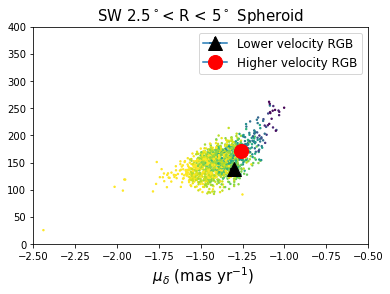}}\\
    \caption{Radial velocity versus pmra and pmdec for the NE (top row), SE (middle row) 3$\rlap{.}^{\circ}$0 -- 5$\rlap{.}^{\circ}$0 and SW (bottom row) 2$\rlap{.}^{\circ}$5 -- 5$\rlap{.}^{\circ}$0 (outer) sub-regions. Simulated data points for the disc components and for the spheroid component of the SMC are shown with distance indicating colour axis and the lower velocity RGB components and higher velocity RGB components are marked with black triangles and red circles, respectively.
    }
    \label{fig:sim_rv_d}
\end{figure*}

\subsection{Comparison with simulations}
\label{section 3.3}
The \textit{N}-body simulations of the Magellanic System by \cite{Diaz2012} modelled the SMC as a multi-component system consisting of a rotating disc, a non-rotating spheroid (the authors considered three models for the spheroidal component) and a dark matter halo and the LMC as a point mass. The SMC model with an extended spheroid (truncation radius = 7.5 kpc and scale-length = 1.5 kpc) best reproduced the observed features. The Milky Way is modelled as a disc, bulge and Navarro-Frenk-White (NFW) dark matter halo \citep{nfw1996}. The panels of Fig. \ref{fig:sim_rv_d} show the comparison of the estimated proper motions ($\mu_\alpha$ and  $\mu_\alpha$ in mas yr$^{-1}$) and the radial velocities (km s$^{-1}$) of the lower velocity RGB component (black point) and higher velocity RGB component (red point) with the corresponding values from the simulated data points (for both the spheroidal and disc components separately) that are coloured according to distance for the NE, SE and SW outer sub-regions. The observed median values of the proper motions and the peak of the radial velocity are taken from Tables \ref{tab:2} and \ref{tab:1} respectively. The distance to the centre of mass of the SMC used by \cite{Diaz2012} in the simulations is 61.6 kpc. As the mean distance to the main body of the SMC estimated from the RC stars by \cite{omkumar} is 65.8 kpc and the proper motion values of the higher velocity RGB component are comparable to those of the main body of RC, we considered the mean distance to the SMC as 65.8 kpc for the simulated points.  
This difference in the distance is applied as a systematic offset to the distance of the simulated data points and their proper motion values are also re-scaled accordingly before comparing with the observed values.
Most of the simulated data points are distributed at the main body distance of the SMC. But there are sub-structures at closer and farther distances from the main body. These sub-structures are suggested to be formed during the tidal interactions between the SMC and the LMC. The sub-structures are more prominent in the disc component. \cite{Diaz2012} suggested that the foreground sub-structure is part of the Magellanic Bridge and the sub-structure at farther distance is referred to as the Counter-Bridge.\\ 
\indent The plots (especially the comparison of the observed values with the predicted values for stars in the disc component) indicate that the two RGB components, separated in radial velocity and proper motion are also separated by line of sight distance. In all sub-regions shown, the lower radial velocity RGB component with higher proper motion values (shown as black point) is at a shorter line of sight distance compared to the higher radial velocity RGB component with lower proper motion values (shown as red point). In the NE and SE outer sub-regions, the higher radial velocity RGB component is at the SMC main body distance and the lower radial velocity RGB component is in the foreground of the main body. This is consistent with the results from the comparison of the proper motion values of the RC and RGB stars in Sect. \ref{section 3.2}. We did not find any signature of the presence of a Counter-Bridge in the NE and SE outer sub-regions from either the RGB or RC stars analyses. The bottom panels of Fig. \ref{fig:sim_rv_d} suggest that in the SW outer sub-region the lower radial velocity RGB component is at the main body distance and the higher radial velocity RGB component is at a farther distance from the main body. 
This indicates the presence of RGB stars in the Counter-Bridge region of the SW SMC. As discussed in Sect. \ref{section 3.2} we did not find any signature of the presence of the Counter-Bridge from the study of RC stars. The RC stars at the farther distance are expected to be fainter than the main body RC stars and that might be the reason for the non-identification of this  feature in \textit{Gaia} data (where we expect the RC magnitudes in the Counter-Bridge to be close to the limiting magnitude). However, this feature was not identified either from the study of VMC data, where the RC magnitudes corresponding to the Counter-Bridge are expected to be $\sim$ 4--5 mag brighter than the limiting magnitude.\\ 
\indent The comparison with the simulations suggests that the lower velocity RGB components identified in the NE and the SE outer sub-regions are at a closer distance to us than the main body of the SMC. Hence they are part of the foreground sub-structure identified by the RC stars in earlier studies. Comparison with the simulations in the SW outer sub-region, indicates the presence of RGB stars in a sub-structure (Counter Bridge) behind the main body of the SMC.
\subsection{Epoch of formation of the RGB sub-structure }
\label{section 3.4}

As RC stars are standard candles, \cite{omkumar} calculated the distances of the foreground and main-body population of the RC stars in the SMC using \textit{Gaia} DR2 data. 
Using data from \textit{Gaia} EDR3 data and following the same analysis as \cite{omkumar}, we found that the foreground population of RC stars in the outer NE sub-region is located  at 11.67 $\pm$ 1.80 kpc in front of the main body RC population of the SMC. We also estimated the distance differences between the dual population in the outer SE sub-region as 13.17 $\pm$ 1.09 kpc. Since the lower and higher velocity components of the RGB stars have similar proper motion values as the foreground and main body RC stars, we assume a similar distance separation between the two radial velocity components of the RGB stars.  Also, the comparison with simulations suggests a similar distance separation between the RGB components. Using the relative velocities between the lower and higher velocity components in the NE (34.41 $\pm$ 3.62 km s$^{-1}$) and in the SE (45.45 $\pm$ 1.67 km s$^{-1}$) and a distance separation of 11.67 $\pm$ 1.80 kpc in the NE and 13.17 $\pm$ 1.09 kpc in the SE between the two populations, we estimated the time of formation of this foreground stellar sub-structure as the difference in the distance divided by the difference in the radial velocity. We found the time scale as 283 $\pm$ 25 Myr and  331 $\pm$ 61 Myr based on the values from the NE and SE respectively. Taking a mean of the two values, we see that the foreground stellar sub-structure was formed  307 $\pm$ 65 Myr ago. This time scale is comparable with the recent direct collision/interaction between the MCs, predicted by simulations (\citealp{Besla2012, Diaz2012}) suggesting that the foreground stellar sub-structure might have formed in the most recent tidal interaction between the MCs.\\ 
\indent This is a first order calculation and we note that we have assumed the difference in distance and velocity only along the line of sight and there could be contributions from the components across the sky plane. As shown in \cite{omkumar}, the foreground RC sub-structure has $\sim$ 35 km s$^{-1}$ slower tangential velocity than the main body of the SMC. Using that we can estimate the resultant velocity difference, but we do not have an independent estimate for the change in position of the sub-structure across the sky plane, which is essential to find the resultant distance difference. Another assumption in the calculation is that the the velocity difference between the main body and the foreground stellar sub-structure remains constant from the the time of formation until today. If the sub-structure is currently accelerating then the estimated time scale is a lower limit and if the sub-structure is currently decelerating then the estimated time scale is an upper limit.  

\section{Summary}
\label{section4}
The eastern region of the 
SMC, in the direction of the Magellanic Bridge, is found to have a foreground stellar sub-structure which is identified as a distance bimodality in the previous studies using RC stars. The RC stars in the sub-structure were also found to be kinematically distinct from RC stars in the main body of the SMC. This sub-structure was suggested to have been formed during the last tidal interaction between the Magellanic Clouds around 300 Myr ago, which formed the MB. If the sub-structure was formed in a tidal interaction then different stellar populations older than 300 Myr are expected to be found in this sub-structure. However, RR Lyrae stars which are older than 10 Gyr are not found in this sub-structure, especially in the NE region of the SMC (see \citealt{muraveva2018}, \citealt{tatton2020}, \citealt{omkumar} for more details). Hence it is essential to look for the presence/absence of other stellar populations in this foreground sub-structure to understand its nature and origin. Interestingly, RGB stars in the eastern region of the SMC show a bimodality in their radial velocity distribution. In this study, we investigate the connection between the bimodality in the radial velocity distribution of RGB stars and the bimodality in the distance distribution of RC stars, observed in the eastern regions of the SMC. \\
\indent The two components in the radial velocity distribution of RGB stars are separated by $\sim$ 34 km s$^{-1}$ and $\sim$ 45 km s$^{-1}$  in the NE and SE regions, respectively. Using archival spectroscopic data and \textit{Gaia} EDR3 astrometric data of RGB stars, we found that the RGB stars with lower radial velocities have higher proper motion values than RGB stars with higher radial velocities. In these regions, the observed proper motion values of the RGB stars having lower and higher radial velocities are comparable with those of the foreground and main body RC stars (separated by a line of sight distance of $\sim$ 11 kpc and $\sim$ 13 kpc in the NE and SE regions respectively) respectively. This suggests that the two RGB populations are separated by a line of sight distance of $\sim$ 11 kpc and $\sim$ 13 kpc in the NE and SE region respectively and the RGB stars with lower radial velocities are part of the foreground stellar sub-structure identified using RC stars. Comparison of the observed properties with \textit{N}-body simulations also suggests that the RGB stars in the lower radial velocity component are at a shorter distance than the main body population. Based on the differences in the distance and radial velocity values, we estimated the time of formation of the foreground sub-structure as 307 $\pm$ 65 Myr ago. This is comparable to the values predicted by simulations for the epoch of the most recent tidal interaction between the Magellanic Clouds. Our results provide evidence for the presence of another intermediate-age stellar population (RGB stars) in the foreground stellar sub-structure of the SMC, which is most likely formed during a tidal interaction about 307 $\pm$ 65 Myr ago. \\
\indent We also identified a bimodal radial velocity distribution of RGB stars in the SW region of the SMC, where a distance bimodality is not found in our analysis of RC stars.  Comparison with the N body simulations of the SMC indicate that the higher radial velocity component in the SW region is at a farther distance than the main body of the SMC. This indicates the presence of a Counter-Bridge, behind the main body of the SMC, in the SW region.\\\\
{\bf Acknowledgements}\\
SS acknowledges support from the Science and Engineering Research Board of India through a Ramanujan Fellowship. 
AOO acknowledges support from the Indian Institute of Astrophysics for carrying out this research. MRC, FN and AOO acknowledge support from the European Research Council (ERC) under the European Horizon 2020 research and innovation programme (grant agreement no. 682115). This research was supported in part by the Australian Research Council Centre of Excellence for All Sky Astrophysics in 3 Dimensions (ASTRO 3D), through project number CE170100013 (RdG). This work has made use of data from the European Space Agency (ESA) mission {\it Gaia} (\url{https://www.cosmos.esa.int/gaia}), processed by the {\it Gaia} Data Processing and Analysis Consortium (DPAC, \url{https://www.cosmos.esa.int/web/gaia/dpac/consortium}). Funding for the DPAC has been provided by national institutions, in particular the institutions participating in the {\it Gaia} Multilateral Agreement. In this work, we have used numpy \citep{numpy}, scipy \citep{virtanen2020}, matplotlib \citep{matplotlib} and astropy,\footnote{http://www.astropy.org} \citep{astropy:2013, astropy:2018}. Finally, it is our pleasure to thank the referee for an encouraging report.\\
\newline
{\bf Data Availability}   
The mean radial velocities and median proper motion values of RGB stars in different sub-regions are provided in various tables in the respective sections of the article. We used the radial velocity data of RGB stars in the SMC which are provided by \cite{dobbie2014} and \cite{deleo2020}. The Gaia data used in the study  were released as part of Gaia Early Data Release 3 (EDR3) and is available in Gaia Archive at \url{https://archives.esac.esa.int/gaia}.


\bibliographystyle{mn2e}
\bibliography{reference_new}


\bsp	
\label{lastpage}
\end{document}